\documentclass[floatfix,showpacs,preprint, amsmath,amssymb,prl]{revtex4}
\usepackage{subfigure}
\usepackage{graphicx}
\usepackage{dcolumn}
\usepackage{bm}
\usepackage[latin1]{inputenc}
\begin{document}

\bibliographystyle{aip}

\title{Magnetism of Covalently Functionalized Carbon Nanotubes}

\author{Elton J. G. Santos, D.~S\'anchez-Portal and A. Ayuela}
\email{eltonjose_gomes@ehu.es}

\affiliation{Donostia International Physics Center (DIPC), Paseo Manuel de Lardizabal 4, 20018 San Sebasti\'an, Spain \\ 
Centro de F\'isica de Materiales, Centro Mixto CSIC-UPV/EHU, Paseo Manuel de Lardizabal 5, 20018 San Sebast\'ian, Spain}

\date{\today}

\begin{abstract}

We investigate the electronic structure of carbon nanotubes functionalized 
by adsorbates anchored with single C-C covalent bonds.  
We find that, despite the particular adsorbate, a spin moment with a universal value of 1.0 $\mu_B$ per molecule
is induced at low coverage. Therefore, we propose a mechanism of bonding-induced magnetism at the carbon surface. 
The adsorption of a single molecule creates a  dispersionless defect state 
at the Fermi energy, which is mainly localized in the carbon wall and presents a 
small contribution from the adsorbate. This universal spin moment is fairly independent of the coverage as long as 
all the molecules occupy the same graphenic sublattice. The magnetic coupling between 
adsorbates is also studied and reveals a key dependence on the graphenic sublattice 
adsorption site.

\end{abstract}


\maketitle


Ferromagnetism in otherwise nonmagnetic materials has been experimentally reported for 
a number of nanoscale systems\cite{Esquinazi03,Esquinazi07,Hernando04}. A very active 
line of research leads to carbon-based materials which is related to the field of spintronics.
Most of these experiments on carbon are related to lattice imperfections
or disorder. Some examples are given by proton-irradiated thin carbon
films\cite{Esquinazi07,Esquinazi03}, nitrogen- and carbon-ion-implanted 
nanodiamond \cite{Ajayan05}, pyrolytic graphite containing a high defect 
concentration\cite{Esquinazi02} or by vacancies created by scanning 
tunneling microscopy in multilayered graphene\cite{Guinea10}.
We propose that despite of these defective systems the well 
known sidewall functionalization\cite{Hirsch02} could also be used to induce 
a magnetic state in single wall carbon nanotubes (SWNT).

In this letter, we show that, when a single C-C covalent bond is established
with a chemisorbed adsorbate at the carbon surface, a spin moment 
is induced in the system. This moment has a universal value of 1.0 $\mu_B$,
independent of the nature of the adsorbate, and we 
show that this effect occurs for a wide class of organic 
and inorganic molecules with different chemical activity (e.g. alkanes, polymers, diazonium 
salts, aryl and alkyl radicals, nucleobases, amido and amino groups, acids).
When several adsorbates are simultaneously adsorbed at the wall, we have found  
that, for metallic or semiconducting SWNTs, only the configurations with all the 
adsorbates in one sublattice  
develop a spin moment. We refer to the two sublattices that define the bipartite 
structure of a graphene layer. Metallic tubes exhibit a ferromagnetic (FM) behaviour, while for semiconducting tubes 
FM and antiferromagnetic (AFM) spin solutions are almost degenerate. 
For molecules at opposite sublattices no magnetic solution can be stabilized in both types of nanotubes.

Our findings are obtained within the density functional theory\cite{Kohn65}
as implemented in the SIESTA code~\cite{Soler02}.
We use the generalized gradient approximation\cite{pbe-functional} and 
Troullier-Martins~\cite{Troullier91} pseudopotentials.
The structures in the periodic supercell
method contain up to 310 atoms\cite{supercell} with adsorbate concentrations 
(defined as the ratio between molecules and the number of atoms in the SWNT) 
ranging from 0.6\% to 25.0\%. 
The atomic coordinates
were relaxed using a conjugated
gradient algorithm until all the force components 
were smaller than 0.04~eV/\AA.
To prevent spurious interactions the 
minimum distance between the walls of
neighboring SWNTs was 18~\AA.
The real-space grid used to calculate the Hartree and
exchange-correlation contribution to the
total energy and Hamiltonian was equivalent
to a 150~Ry plane-wave cutoff. The  $k$-point sampling was
equivalent to a 1$\times$1$\times$136 sampling~\cite{MonkhorstPack76}
of the Brillouin zone of a single tube cell.
We have done some calculations using the VASP code~\cite{Kresse93,Kresse96}. We used 
projected-augmented-wave potentials with a well converged plane-wave cutoff of 400~eV.  
The  rest  of  computational details  was  fixed  
as in  the  SIESTA method. The results
obtained with VASP are almost identical to those obtained with SIESTA.

\begin{figure}
\includegraphics[width=3.3500in]{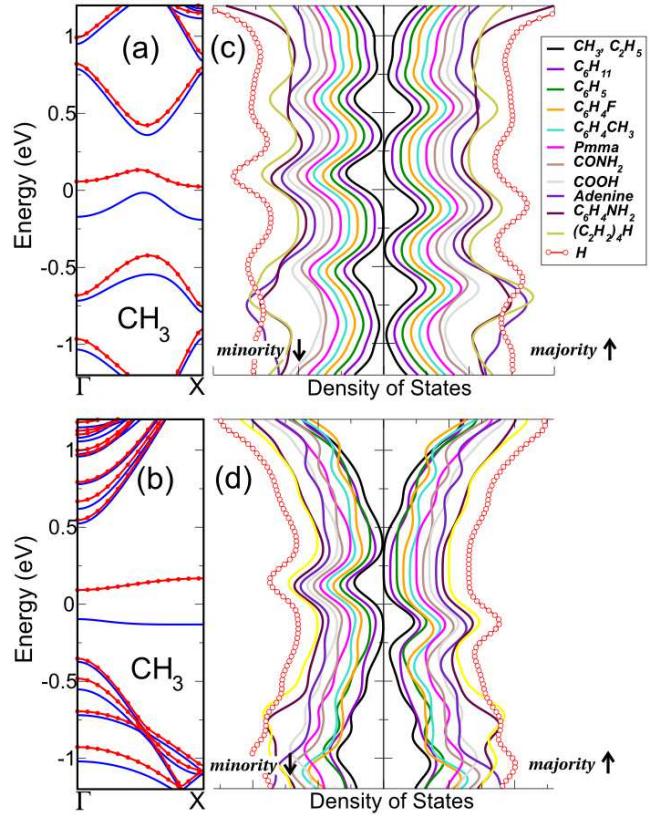}
\caption{(Color online) Spin polarized band structure and density of states for 
(a)-(b) (5,5) and (c)-(d) (10,0) SWNTs with a single adsorbate (per supercell) of different types 
chemisorbed to a carbon atom through a single C-C covalent bond. In panels (a) and (b), 
the blue (dark) and red (bright) lines denote the majority and minority spin bands, respectively. 
For clarity, the curves in panel (c) and (d) have been shifted and smoothed with a Lorentzian 
broadening of 0.12 eV. The Fermi energy is set to zero in all panels.
}
\label{fig1}
\end{figure}

In order to understand the origin of a common spin moment when a covalent bonding
is attached to the tube wall, the spin polarized band structure of a CH$_3$
molecule chemisorbed on top of a C atom is shown in Figure \ref{fig1} 
for (a) (5,5) and (b) (10,0) SWNTs. 
In both cases, a defect state appears pinned at the 
Fermi level (E$_{F}$) with full spin polarization. This state 
is mainly composed by $p_z$ orbitals of the C neighbors to the
saturated site, with almost no contribution from the adsorbate. 
In fact, a detailed Mulliken analysis of this 
$p_z$-defect state assigns a small contribution of the spin moment to the adsorbate. 
This indicates that the adsorbate has a primary role in creating the bond with the nanotube, 
and the associated defect level, but it does not appreciably contribute to the spin moment.
More complex adsorbates, notwithstanding of the biological and chemical activity 
(e.g. alkanes, polymers, diazonium salts, aryl and alkyl radicals, nucleobases,
amido and amino groups, acids), show a 
similar behaviour. This is observed 
in the density of states (DOS) per spin channel for metallic (5,5) and 
semiconducting (10,0) SWNTs shown in Figure \ref{fig1}(c) and 
\ref{fig1}(d), respectively. Several common points are worth mentioning:
(i) All molecules induce a spin moment of 1.0 $\mu_{B}$ 
localized at the carbon surface;
(ii) The origin of the spin polarization corresponds 
to the $p_{z}$-defect state as explained above for the CH$_3$ molecule;
(iii) The DOS around E$_{F}$ follows the same pattern in all cases. 
This match demonstrates that the spin moment 
induced by the covalent functionalization is independent of 
the particular type of adsorbate. These results also point out the complete analogy 
between a single C-H bonding with more complex C-C arrangements, which is not an obvious behaviour.

\begin{figure}
\includegraphics[width=3.900in]{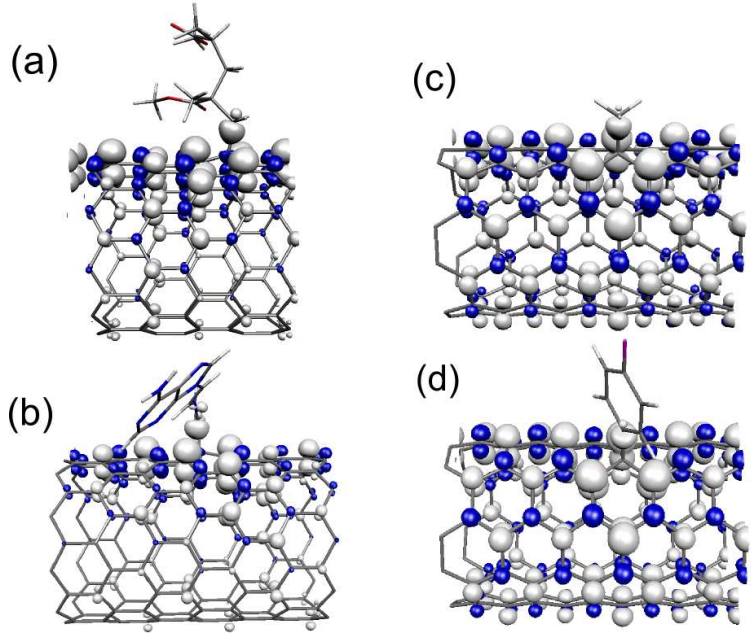}
\caption{(Color online)  Isosurface for the magnetization density induced by some adsorbates at the  
SWNT surface: (a) Pmma and (b) Adenine group in a (10,0); and (c) CH$_{3}$ and (d) C$_{6}$H$_{4}$F in a (5,5).
Majority and minority spin densities correspond respectively to light and  dark  surfaces, which 
alternate  on the honeycomb lattice with long decaying order in all cases. 
The cutoff is at $\pm$0.0133  $e^-/bohr^3$.
}
\label{fig2}
\end{figure}

Next we study the spin polarization texture induced by the adsorbates
on the carbon nanotube wall. The analysis of local magnetic moments for all the adsorbates 
assigns general trends to both SWNTs. The C atoms that participate directly in the 
bond formation, at either the molecule or the surface, show a local spin moment smaller 
than $\sim 0.10 \mu_{B}$. However, the wall carbon atoms contribute with 0.40 $\mu_{B}$ 
in the three first C nearest-neighbors, -0.10 $\mu_{B}$ 
in the next nearest-neighbors, 0.20 $\mu_{B}$ in the third-neighbors.
The adsorbate removes a $p_{z}$ electron from the adsorption site, and leave the $p_z$ states of the 
nearest carbon neighbours uncoordinated and localized. This gives rise to a 
defect state localized in the carbon layer and reminiscent of that 
of a vacancy in a $\pi$-tight-binding model of graphenic nanostructures.
The carbon spins polarize parallel 
(antiparallel) respect to the C atom that binds to the surface 
when sitting in the opposite (same) sublattice.  
Figure \ref{fig2} shows the magnetization density
in semiconducting (10,0) and metallic (5,5) SWNTs for several molecules:
(a) Pmma polymer chain\cite{Fisher00}, (b) Adenine group nucleobase\cite{Bianco09}, (c) CH$_3$ molecule\cite{Smalley02} 
and (d) C$_6$H$_4$F salt\cite{Tour01}. The spin density in the metallic (5,5) (Figure \ref{fig2}(c) and \ref{fig2}(d)) 
is more spread over the whole surface than in the semiconducting (10,0) (Figure \ref{fig2}(a) and \ref{fig2}(b)). 
This indicates that electronic character of the nanotube wall plays a role 
in mediating the interaction between adsorbates.

\begin{figure}
\includegraphics[width=3.400in]{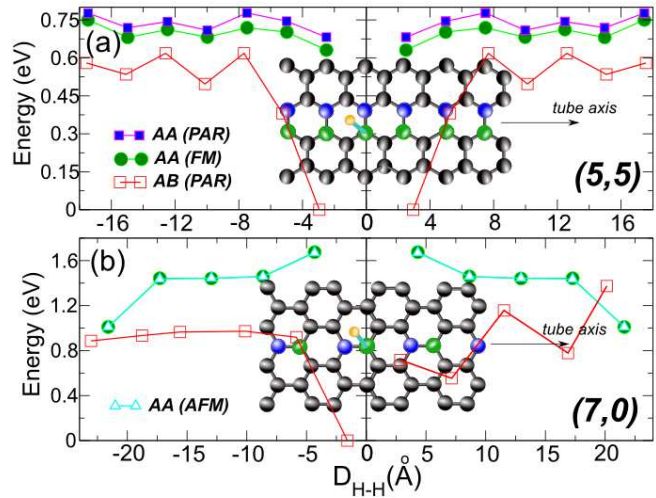}
\caption{(Color online) Variation of total energy with the H positions
for the distinct magnetic solutions in the two graphitic sublattices (AA and AB) for (a) (5,5) and (b) (7,0)
SWNTs. The empty and filled squares correspond to PAR
spin solutions in AB and AA sublattices, respectively.
The circles and triangles indicate the FM and AFM 
solutions, respectively, at the same sublattice.
}
\label{fig3}
\end{figure}

Now we address the energy stability of the different magnetic solutions 
when two molecules are adsorbed.
We focus on a chemisorbed molecule at the nanotube surface by looking at H as an example. 
For the metallic (5,5) and semiconducting (7,0) SWNTs, we calculate 
the variation of the total energy for several spin alignments 
as a function of the distance between the adsorbates at large dilution ($\sim$0.6\% adsorbate concentration). 
The used geometry along tubes is  shown in the insets
of Figure \ref{fig3}(a) and \ref{fig3}(b). 
One H  is  sited  at the  origin; 
and another, in  different positions along of the
tube axis (see background pictures).
Several observations can be first made on the stability when
two adsorbates are located  at the  same sublattice (AA configurations).
In the metallic (5,5), the FM configuration is most stable than the non-magnetic one (PAR). 
The energy difference between these two spin solutions along the tube axis oscillates 
and no AFM solution could be stabilized at all. 
In the semiconducting (7,0), the FM and AFM solutions are almost degenerate, with a small 
energy difference (exchange coupling).

If the two molecules are now located at different sublattices 
(AB configurations), we were not able to stabilize any 
magnetic solution for both nanotubes.
Instead the systems is more stable without a local spin moment. 
This behaviour for adsorbates at opposite sublattices can be traced back 
to the interaction between the defect levels.  While 
for AA configurations the interaction is negligible, for AB ones 
this interaction opens a bonding-antibonding
gap around $E_{F}$ in the $p_z$ defect band and, thus, contributes to the 
stabilization of PAR solutions.
If the gap is larger than the spin splitting of the majority 
and minority spin defect bands the system will be non-magnetic\cite{Co-paper10, Palacios08}. 
In fact, our detailed analysis of the band structure  
fully confirmed this explanation. However, it is worth noting that AB adsorption 
seems to be always more stable in our calculations. This indicates that if the adsorption 
takes place at random sites, the magnetic solutions will only be stable for 
low density functionalization. On the other hand, calculations at high adsorbate coverage (from 12.5\% to 25\% concentration) show 
that for adsorbates in the same sublattice, the system stabilizes the magnetic solutions. 
The interaction between molecules remains quite small and they generate a spin moment 
of $\sim$1.0 $\mu_{B}$ per molecule independently of the coverage. For even higher concentrations, 
$\sim$50.0 \%, the chemisorbed molecules are not structurally stable, and  half of them move away from the surface.


%


%

In summary, we have shown that
sidewall covalent functionalization creates new routes to achieve magnetism in 
carbon nanotubes. Despite the adsorbate, and its 
chemical or biological activity, a spin moment with a value of 
1.00 $\mu_{B}$ is induced in the nanotubes 
when the molecule is attached through a single C-C bond. 
We find that adsorbates at the same sublattice order magnetically. 
For adsorbates at the different sublattices, 
their strong interaction prevents the formation of local spin moment.

We acknowledge support from Basque
Departamento de Educacion and the UPV/EHU (Grant
No. IT-366-07), the Spanish Ministerio de Educaci\'on
y Ciencia (Grant No. FIS2010-19609-CO2-02) and the
ETORTEK program funded by the Basque Departamento de 
Industria and the Diputacion Foral de Guipuzcoa.

\bibliographystyle{apsrev}

\end{document}